\documentclass[aps,prd,nofootinbib,preprint]{revtex4-1}
\usepackage{graphicx}
\usepackage{xcolor}
\usepackage{amsmath}

\def\be{\begin{equation}}
\def\ee{\end{equation}}
\def\bea{\begin{eqnarray}}
\def\eea{\end{eqnarray}}

\begin{document}
\title{An Elusive Vector Dark Matter}

\author{\vspace{1cm} Chuan-Ren Chen,  Yu-Kuang Chu and  Ho-Chin Tsai }

\affiliation{
\vspace*{.5cm}
\mbox{Department of Physics, National Taiwan Normal University, Taipei 116, Taiwan}\\
\vspace*{1.5cm}}
\begin{abstract}
Even though the sensitivity of direct dark matter search experiments reach the level about $10^{-45}~{\rm cm}^2$, there is no confident signal of dark matter been observed. 
We point out that, if dark matter is a vector boson, the null result in direct dark matter search experiments may due to the destructive effects in dark-matter-nucleon elastic scattering. 
We illustrate the scenario using a modified Higgs portal model that includes exotic quarks. 
The significant cancellation can occur for certain mass gap between heavy quark and dark matter.
As a result, the spin-independent dark-matter-nucleon elastic scattering is so suppressed that the future direct search experiments can hardly observe the signal of dark matter.

\end{abstract}

\maketitle

\section{Introduction}
\label{sec: intro}
The current dark matter relic abundance in our Universe has been measured by WMAP \cite{Hinshaw:2012aka} and recently by Planck \cite{Ade:2013zuv} with the combined value
\be
\Omega_{DM}h^2 = 0.1199 \pm 0.0027.
\label{eq:planck}
\ee 
However, we have very little knowledge about dark matter. 
The nature of  dark matter particle is one of the most challenging problems in particle physics. The most attracting candidate is the weakly interacting massive particle (WIMP). Certainly, Standard Model (SM) in particle physics has no proper candidate. There are many proposals beyond the SM such as the lightest neutralino in supersymmetry~\cite{Jungman:1995df}.  

The searches of dark matter can be categorized into three categories: collider experiment, which looks for the signal of missing  transverse momentum that is contributed by dark matter produced from the collision of two SM particles; direct detection experiment, which searches for the scattering of dark matter off atomic nuclei; indirect detection, which looks for the products of dark matter annihilation in our Universe. Recently, disagreements between astrophysical background and observation that may be the hint of dark matter have been observed in cosmic gamma-ray~\cite{Su:2010qj,Hooper:2010mq} and positron data~\cite{Adriani:2008zr, Aguilar:2013qda}. But direct detection is required to show the existence of dark matter.

Null results from the direct search constrain the scattering cross section between dark matter and nucleus. The current upper bound of WIMP-nucleon scattering cross section about $10^{-45}~\rm{cm}^2$ is set  by LUX experiment~\cite{Akerib:2013tjd}.   
Therefore,  the  crucial question we might ask is why the scattering cross section is so tiny that  these sophisticated detectors are incapable of the detection of dark matter. 
It may be simply because  the mass of dark matter is not within the sensitive region of these detectors or interaction between dark matter and nucleon is extremely small. 
 In this paper, we point out that, for a vector dark matter particle, cancellation between Feynman diagrams can naturally happen. As a result, we are able to realize the tiny spin-independent elastic scattering cross section between dark matter and nucleon.  
For illustration, we study a simple model in which the dark matter candidate is a SM singlet spin-1 gauge boson associated with  $U(1)_{X}$ in dark sector. When heavy quarks  are included, the scattering cross section can be highly suppressed.  

The rest of paper is organized as follows. We begin in Sec.~\ref{sec: model} with an introduction of the model.   In Sec.~\ref{sec:xs}, we calculate the elastic scattering cross section and show the cancellation between diagrams. Our conclusion appears in Sec.~\ref{sec: con}.  

\section{Vector Dark Matter}
\label{sec: model}

Spin-1 vector dark matter appears in  many popular models, such as Kaluza-Klein photon in universal extra dimension~\cite{Cheng:2002ej,Servant:2002aq,Servant:2002hb,Arrenberg:2008wy} and T-odd photon in Little Higgs model with T-parity~\cite{Cheng:2003ju,Cheng:2004yc,Low:2004xc,Hubisz:2004ft,Birkedal:2006fz,Asano:2006nr}. Here we consider a simple model that includes dark matter interactions to quarks in the Higgs portal model. 

Dark matter particle is a vector boson associated with gauge symmetry $U(1)_X$~\cite{Kanemura:2010sh,Lebedev:2011iq,Djouadi:2011aa,Farzan:2012hh,Baek:2012se,Yu:2014pra}. The gauge invariant Lagrangian can be written as
\footnote{We neglect the kinetic mixing term $X_{\mu\nu}B^{\mu\nu}$ by assuming it is extremely small, and therefore,  it  does not affect our study.}
\be
{\cal L}_{VDM} = -\frac{1}{4}X_{\mu\nu}X^{\mu\nu}+\frac{1}{2}M_{X}^2  X_\mu X^\mu +\frac{1}{4}\lambda_X(X_\mu X^\mu)^2 +\frac{1}{2}\lambda_{XH}X_\mu X^\mu H^\dagger H,
\label{eq:hvv}
\ee
where field strength tensor $X_{\mu\nu}=\partial_\mu X_\nu-\partial_\nu X_\mu$, $X_\mu$ is dark matter field, $M_X$ is mass of dark matter particle and $H$ is SM Higgs field.  
The last term describes the interaction between dark matter and the SM Higgs boson and contributes to mass of dark matter after electroweak symmetry is broken. 
The mass of vector dark matter is given as $m_X^2=M_{X}^2+\lambda_{XH} v^2/2$, where $v$ is the Higgs VEV.
Two SM $SU(2)_L$ singlet right-handed quark fields $q_1$ and $q_2$ are introduced. The Lagrangian is then given as
\bea
{\cal L}_q &=& \bar{q}_1 i \gamma_\mu(\partial^\mu -ig_1 Y_1^{q_1} B^\mu-ig_X Y_X^{q_1} X^\mu) q_1\nonumber\\
                   & & + \bar{q}_2 i \gamma_\mu(\partial^\mu -ig_1 Y_1^{q_2} B^\mu-ig_X Y_X^{q_2} X^\mu) q_2,
\label{eq:quark1}
\eea
where $g_{1}$ and $B^\mu$ are the gauge coupling strength and gauge field of SM $U(1)_Y$, respectively; $g_X$ is the gauge coupling strength of $U(1)_X$, 
$Y^{q_{1(2)}}_1$ is the hypercharge of $q_{1(2)}$ under $U(1)_Y$, and $Y^{q_{1(2)}}_{X}$ is  $U(1)_X$ charge for $q_{1(2)}$. 
We can transform $q_1$ and $q_2$ to the right-handed SM  quark and heavy exotic quark $q$ and $q_H$ as
\be
q_{R} = \frac{q_1 + q_2}{\sqrt{2}}~~{\rm and}~~q_{H_R} = \frac{q_1-q_2}{\sqrt{2}}.
\ee
With $Y^{q_1}_1=Y^{q_2}_1$ and $Y^{q_1}_1+Y^{q_2}_1=Y_q$ that is the SM hypercharge of quark $q_R$, and $Y^{q_2}_X=-Y^{q_1}_X=-Y^\prime_q$, we have
\bea
{\cal L}_q &=& \bar{q}_R i \gamma_\mu(\partial^\mu -i\frac{g_1}{2}Y B^\mu) q_R+ \bar{q}_{H_R} i \gamma_\mu(\partial^\mu -i\frac{g_1}{2}Y B^\mu) q_{H_R} \nonumber\\
                   & & + g_X Y^\prime_q (\bar{q}\gamma_\mu X^\mu P_R q_H + h.c.).
\label{eq;quark2}                   
\eea
The last term of Eq.~(\ref{eq;quark2}) gives the interaction between dark matter and SM quark with coupling strength $g_X Y^\prime_q$. 
For the mass of $q_H$, we assume there exists a left-handed $q_{H_L}$  to form a Dirac mass term $m_{q_H} \bar{q}_Hq_H$. 
The parameters relative to the calculation below are $m_X$, $m_{q_H}$, $\lambda_{XH}$, $g_X$ and $Y_q^\prime$.

\section{Elastic Scattering Cross Section}
\label{sec:xs}

\begin{figure}[htbp]
\begin{center}
\includegraphics[scale=0.65,clip]{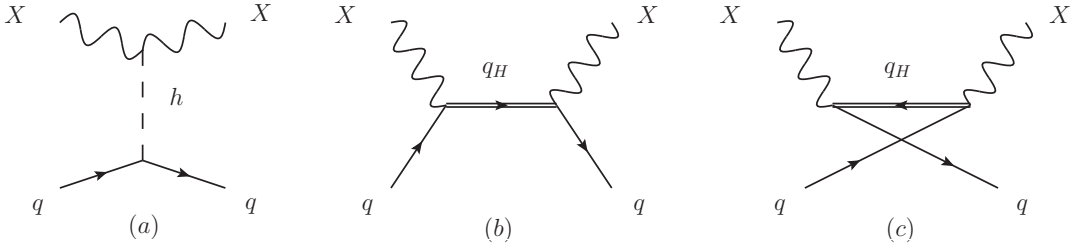}
\caption{Three Feynman diagrams for vector dark matter scattering with quark: (a) t-channel Higgs mediated, (b) s-channel heavy quark mediated and (c) u-channel heavy quark mediated. }
\label{fig:FD}
\end{center}
\end{figure}

Elastic scattering between dark matter $X$ and quark inside the nucleon involves  three diagrams as shown in Fig.\ref{fig:FD}. 

Following calculations in~\cite{Servant:2002hb}, we calculate the dark-matter-nucleon scattering amplitudes $M_h$ and $M_{q_{H}}$ for the Higgs-mediated (Fig.~\ref{fig:FD}(a)) and heavy-quark-mediated (Fig.~\ref{fig:FD}(b) and Fig.~\ref{fig:FD}(c)) diagrams, respectively.  
In non-relativistic limit, we have
\bea
{\cal M}_h&=& i \lambda_{XH} m_q \bar q(p_3) \left[ \epsilon_\mu^*(p_4). \epsilon^\mu(p_1)   \frac{1}{(p_1-p_4)^2-m_{h}^2} \right] q(p_2) \nonumber\\
&\sim& -i \epsilon_{\mu}^*(p_4)\epsilon_{\nu}(p_1) (\gamma_q) g^{\mu\nu} \bar q(p_3)  q(p_2)\\
{\cal M}_{q_H}&=& -i g_X^2 Y_q^{\prime 2}\bar q(p_3) [ \epsilon^*_\mu(p_4)\gamma^\mu  P_R \frac{\not\!p_1+\not\!p_2+m_{X}}{(p_1+p_2)^2-m_{X}^2} \epsilon_\nu(p_1)\gamma^\nu P_R   \nonumber\\
&& +   \epsilon_\nu(p_1)\gamma^\nu  P_R \frac{\not\!p_2-\not\!p_4+m_{X}}{(p_2-p_4)^2-m_{X}^2}\epsilon_\mu^*(p_4)\gamma^\mu  P_R ] q(p_2) \nonumber\\
&\sim&-i g_X^2 Y_q^{\prime 2} \epsilon_{\mu}^*(p_4) \epsilon_{\nu}(p_1)  \bar q(p_3) \left[ -S_q E^{\mu\nu} + A_q \tilde E^{\mu\nu} \right] P_R q(p_2)
\eea 
where 
\bea
E^{\mu\nu}=\gamma^\mu \gamma^0 \gamma^\nu + \gamma^\nu \gamma^0 \gamma^\mu\;,\;\;\; 
\tilde E^{\mu\nu}=\gamma^\mu \gamma^0 \gamma^\nu - \gamma^\nu \gamma^0 \gamma^\mu=2 i \epsilon^{0\mu\nu\rho}\gamma_\rho \gamma_5.
\eea 
The coefficients $\gamma_q$, $S_q$ and $A_q$ can be derived as  
\bea
\gamma_q=\lambda_{XH} \frac{m_q}{m_{h}^2}\;,\;\;\;\;\; S_q=g_X^2 Y_q^{\prime 2} \frac{E_q (m_{X}^2+m_{qH}^2)}{(m_{X}^2-m_{qH}^2)^2}\;,\;\;\;\;\;A_q=g_X^2 Y_q^{\prime 2} \frac{m_{X}}{m_{X}^2-m_{qH}^2}\;. 
\label{eq:sq}
\eea
Note that $\gamma_q$ and $S_q$ contribute to so-called spin-independent (SI) cross section, while $A_q$ is related to spin-dependent (SD) cross section.

In the extreme non-relativistic limit, the elastic scattering cross section between dark matter and nuclear can be divided into two cases: scalar interaction and spin-spin interaction.  The  "standard" total cross section at zero momentum transfer $\sigma_0^{scalar}$ and $\sigma_0^{spin}$~\cite{Jungman:1995df} can be obtained as
\bea
&&\sigma_0^{scalar}=\frac{m_N^2}{4\pi(m_{X}+m_N)^2}(Z f_{p} + (A-Z)f_{n})^2\\
&&\sigma_0^{spin}=\frac{2}{3\pi}\frac{m_N^2}{(m_{X}+m_N)^2}J(J+1)\Lambda^2
\eea
where $m_N$ is the mass of unclear, $Z$ and $A$ are, respectively, nuclear charge and atomic number, while $f_{p(n)}$ is the dark matter effective  scalar coupling to proton (neutron) and can be expressed as
\be
f_{p(n)} = m_{p (n)} \sum_q \frac{\gamma_q+S_q}{m_q}f_{T_q}^{p,n}.
\label{eq:fpn}
\ee
Numerically, we adopt $f^p_{T_u}=0.023$, $f^p_{T_d}=0.034$, $f^n_{T_u}=0.019$, $f^n_{T_d}=0.041$ and $f^p_{T_s}=f^n_{T_s}=0.14$~\cite{Gasser:1990ce}. 
The  contribution of gluon content of the nucleon is  included in $\gamma_q$~\cite{Jungman:1995df,Servant:2002hb}. 

For spin-spin interaction term, 
\be
\Lambda=\frac{a_p \langle S_p\rangle + a_n \langle S_n\rangle }{J}\;,\;\;\;\; a_{p(n)}=\sum_{q=u,d,s}A_q \Delta^{p(n)}_q\;,
\ee
where $J$ is nuclear spin, $a_{p(n)}$ is effective spin-spin interaction of dark matter and proton (neutron), $\langle S_{p(n)}\rangle /J$ is the fraction of the total nuclear spin $J$ carried by protons (neutrons).
We take $\Delta^p_u=\Delta^n_d=0.78$, $\Delta^p_d=\Delta^n_u=-0.48$ and $\Delta^p_s=\Delta^n_s=-0.15$~\cite{Mallot:1999qb}.
The one nucleon normalized spin-independent and spin-dependent scattering cross section to be compared to the experimental results are 
\bea
&&\sigma_{p(n)}^{SI}=\sigma_{p(n)}^{scalar}=\frac{1}{4\pi}\frac{m_{p(n)}^2}{m_X^2 A^2}(Z f_{p} + (A-Z)f_{n})^2;\label{eq:SI}\\
&&\sigma_{p(n)}^{SD}=\sigma_{p(n)}^{spin}=\frac{1}{2\pi}\frac{m_{p(n)}^2  a^2_{p(n)}}{(m_X+m_{p(n)})^2}.
\label{eq:SD}
\eea
%
\begin{figure}[htbp]
\begin{center}
\includegraphics[scale=0.85,clip]{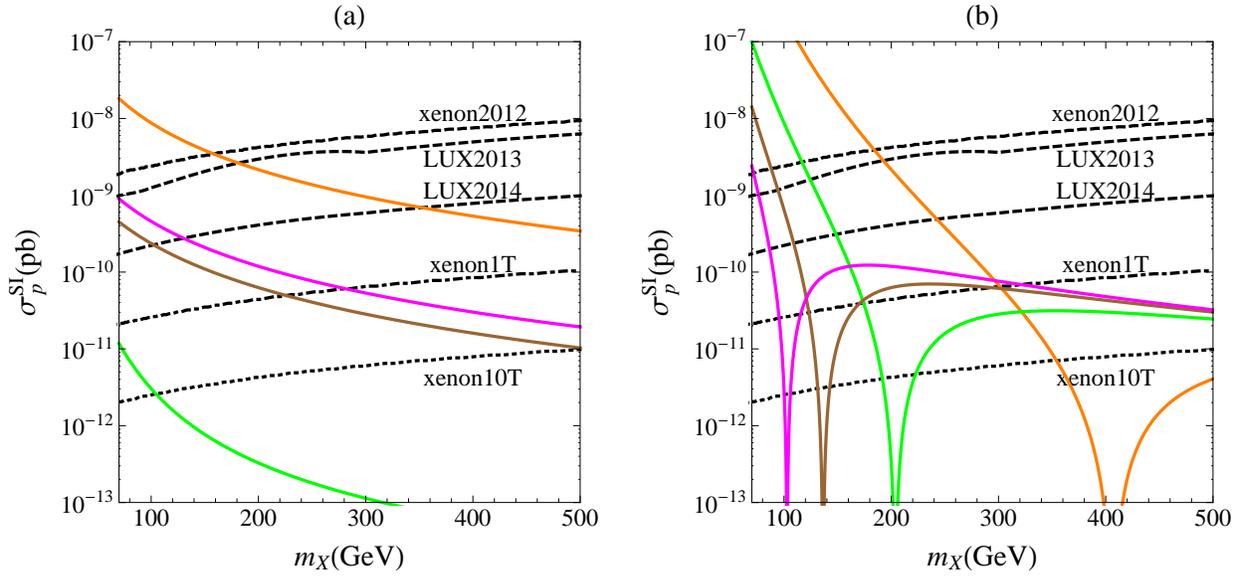}
\caption{The one-nucleon-normalized spin-independent elastic cross section of the vector dark matter $X$ scattering off the proton  is plotted together with the current  experimental limits of XENON100 (2012)~\cite{Aprile:2012nq} and LUX(2013)~\cite{Akerib:2013tjd}. The projected LUX(2014), XENON 1T\cite{Aprile:2012zx}, and XENON10T~\cite{xenon10T} are also shown. The orange, green, brown and magenta color lines refer to $m_{q_H}=m_{X}+$ $20$, $40$, $60$, $80$ GeV in the left panel (a) and $m_{q_H}/m_X=1.1$, $1.20$, $1.30$, $1.40 $  in the right panel (b), respectively.  $\lambda_{XH}=-0.05$ and $\lambda_{q_H}=0.1$ are used.}
\label{fig:dir1_1}
\end{center}
\end{figure}
\begin{figure}[htbp]
\begin{center}
\includegraphics[scale=0.85,clip]{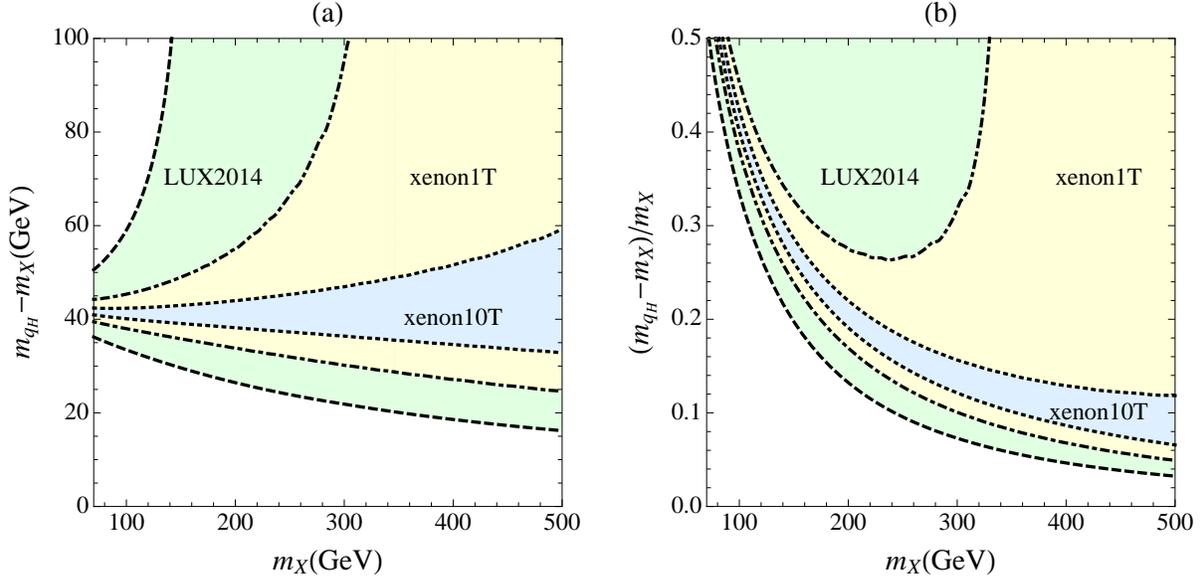}
\caption{ The contour plot for parameter space in which the SI  scattering cross section is below the projected limits of the future experiments.  The green, yellow and blue regions are for LUX(2014), XENON 1T~\cite{Aprile:2012zx}, and XENON10T~\cite{xenon10T}, respectively. (a) ($m_{q_H}-m_X$) v.s. $m_X$; (b) $(m_{q_H}-m_X)/m_X$ v.s. $m_X$.
 $\lambda_{XH}=-0.05$ and $\lambda_{q_H}=0.1$ are used.}
\label{fig:dir1_2}
\end{center}
\end{figure}
We notice that the spin-independent cross section relies on $f_{p(n)}$ in Eq.~(\ref{eq:fpn}). Therefore, $\sigma_{p(n)}^{SI}$ could be far below the sensitivities of current experiments if $f_{p(n)}$ is tiny.  
It is possible to  naturally make effective coupling $f_{p(n)}$ small if there is a destructive effect between $\gamma_q$ and $S_q$ in Eq.~(\ref{eq:fpn}). Such a destruction can be achieved when the sign of the coupling $\lambda_{XH}$  is negative. Meaning that the mass of dark matter shifts to a smaller value after the  electroweak symmetry is broken.  

 In Fig.~\ref{fig:dir1_1}, we show the SI elastic scattering cross section between dark matter $X$ and proton, compared with current limits from XENON100~\cite{Aprile:2012nq} and LUX 2013~\cite{Akerib:2013tjd}. The projected sensitivities of LUX and XENON  experiments in the future are also shown. Here, we set  the parameter $\lambda_{qH}= g_X Y^\prime$ with universal $Y^\prime_q=Y^\prime$ for simplicity. We can see in the left panel that, with $\lambda_{XH}=-0.05$ and $\lambda_{qH}=0.1$ for illustration, the cross section is well below the current limit. In the right panel of Fig.~\ref{fig:dir1_1}, the significant cancellations between $\gamma_q$ and $S_q$ can be seen.  
The mass of dark matter at which the exact cancellation occurs  shifts to a smaller value when the heavy exotic quark mass to dark mater mass ratio $r=m_{q_H}/m_X$ gets larger. This behavior can be easily understood as follows. Since $(1+r^2)/(1-r^2)^2$ in Eq.~(\ref{eq:sq}) is a decreasing function for $r>1$, a smaller $m_X$ is then required for a complete cancellation when heavy quark $q_H$ is heavier (i.e. $m_{q_H}/m_X$ is larger) 

The contour in Fig.~\ref{fig:dir1_2} shows the parameter space where the SI cross section is below the projected sensitivities of upcoming LUX and future XENON experiments.  We show that, when the heavy exotic quark is heavier than the vector dark matter within a certain range, the scalar interaction of dark matter and nuclear can be suppressed significantly. As a result, it is extremely challenging to detect the dark matter, even for the detectors with high sensitivity in the future experiments. For illustration, we adopt the benchmark couplings $\lambda_{XH}=-0.05$ and $\lambda_{q_H}=0.1$. As seen in Fig.~\ref{fig:dir1_2}a, for dark matter mass from  $100$ GeV to $500$ GeV, the $\sigma_{p(n)}^{SI}$ is below the value that can be detected by  the future XENON 10T experiment, if the mass difference between heavy exotic quark and dark matter ($\Delta m$) is about $35~{\rm GeV}\sim 55~{\rm GeV}$ (or about $8\%\sim 50\%$ of $m_X$ shown in Fig.~\ref{fig:dir1_2}b).   
The feature is that the heavier dark matter needs a smaller value of $\Delta m/m_X$ for a complete cancellation.

\begin{figure}[htbp]
\begin{center}
\includegraphics[scale=0.8,clip]{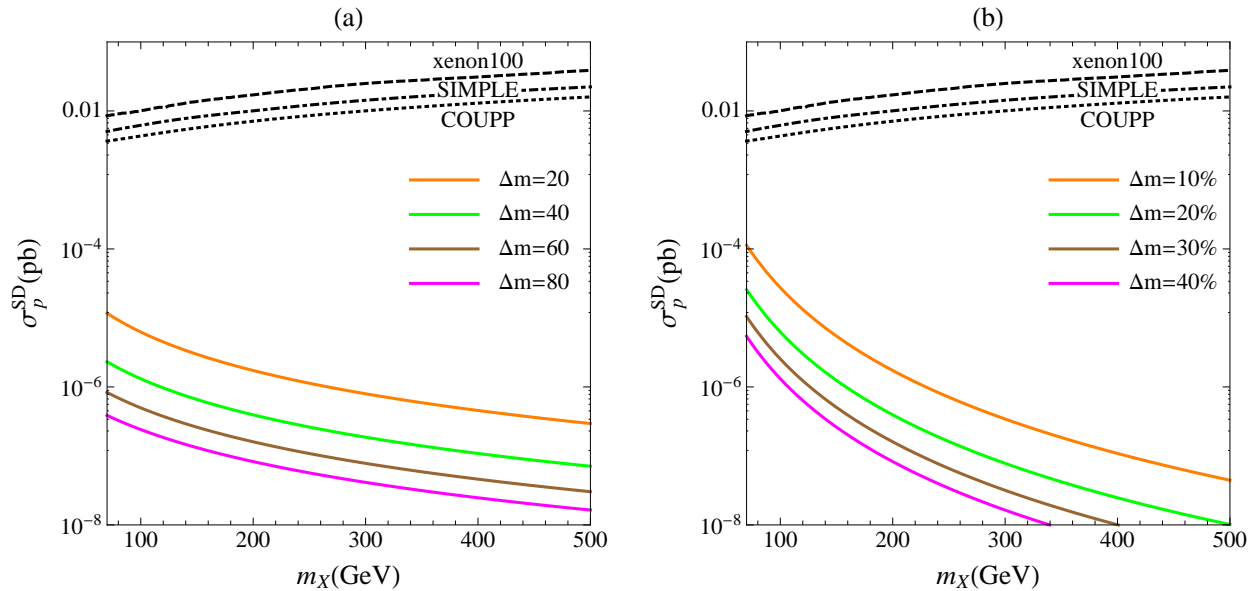}
\caption{The spin-dependent dark-matter-proton cross section, along with current constraints from SIMPLE~\cite{SIMPLE}, COUPP~\cite{COUPP} and XENON 100 \cite{Aprile:2013doa} experiments. The orange, green, brown, magenta solid lines are for (a) $\Delta m=m_{q_H}-m_X=20,~40,~60,~80$ GeV; (b)  $\Delta m/m_X=(m_{q_H}-m_X)/m_X=10\%,~20\%,~30\%,~40\%$.  We use $\lambda_{q_H}=0.1$ here.}
\label{fig:SDp}
\end{center}
\end{figure}

\begin{figure}[htbp]
\begin{center}
\includegraphics[scale=0.8,clip]{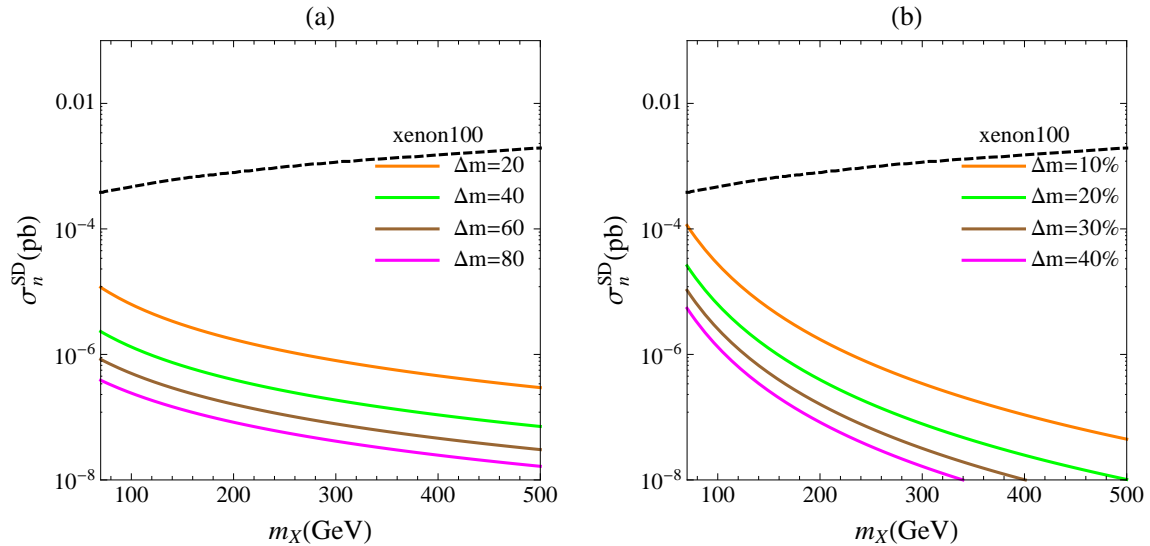}
\caption{The spin-dependent dark-matter-neutron cross section, along with current constraints from  XENON 100 \cite{Aprile:2013doa} experiment. The orange, green, brown, magenta solid lines are for (a) $\Delta m=m_{q_H}-m_X=20,~40,~60,~80$ GeV; (b)  $\Delta m/m_X=(m_{q_H}-m_X)/m_X=10\%,~20\%,~30\%,~40\%$.  We use $\lambda_{q_H}=0.1$ here.}
\label{fig:SDn}
\end{center}
\end{figure}

For spin-dependent cross section, the results are shown in Fig.~\ref{fig:SDp} and Fig.~\ref{fig:SDn} for the proton and neutron, respectively, along with current limits. The constraint for neutron is more stringent than that for proton. However, the constraints are still quite loose. The left panel is for fixed $\Delta m=m_{q_H}-m_X$, while the right panel is for  $\Delta m = (m_{q_H}-m_X)/m_X$. With the parameter $\lambda_{qH}\sim 0.1$, the cross sections for both proton and neutron are well below the bound.  
\section{Discussion and Conclusion}
\label{sec: con}
The sensitivity of dark matter direct search has reached down to the level of $10^{-9}~pb$ for dark-matter-nucleon spin-independent elastic scattering cross section. However, there is no positive signal observed. 
In this paper, we point out the possibility that the null result in direct dark matter search experiments may due to the destructive effects in dark-matter-nucleon scattering. 
We use a simple vector dark matter model for illustration. The spin-1 vector dark matter particle communicates with SM via the Higgs boson and the right-handed heavy exotic quarks. 
The effective dark-matter-nucleon scalar interaction can be highly suppressed because of
the cancellation between the Higgs boson exchange diagram and the diagrams with heavy exotic quark.

Our results show that when the mass difference between the heavy exotic quark and dark mater ($\Delta m = m_{q_H}-m_X$) is within a certain range, the reduction in SI cross section is so significant that even the future XENON 10T experiment can hardly observe the signal of dark matter. 
For a heavier dark matter particle,  the ratio $\Delta m /m_X$ should be smaller for the significant cancellation to occur. 
We also calculate the SD cross sections,  which could constrain the coupling of dark matter to quarks. However, the current limit is still quite loose, therefore the constraints from both neutron and proton data are not stringent. 

Finally, we  comment on the relic abundance. In our scenario, the vector dark matter annihilation processes are similar to the T-odd photon  in the Littlest Higgs model with T-parity. It is shown~\cite{Birkedal:2006fz} that the T-odd photon can nicely explain the relic abundance. Therefore, the vector dark matter in our study will satisfy the observation of relic abundance as well.

\begin{acknowledgments}~This work is supported in part by the National Science Council under Grant No.~NSC 102-2112-M-003-001-MY3. 

\end{acknowledgments}


\begin{thebibliography}{99}


\bibitem{Hinshaw:2012aka} 
  G.~Hinshaw {\it et al.}  [WMAP Collaboration],
  Astrophys.\ J.\ Suppl.\  {\bf 208}, 19 (2013)
  [arXiv:1212.5226 [astro-ph.CO]].

\bibitem{Ade:2013zuv} 
  P.~A.~R.~Ade {\it et al.}  [Planck Collaboration],
  arXiv:1303.5076 [astro-ph.CO].

\bibitem{Jungman:1995df} 
  G.~Jungman, M.~Kamionkowski and K.~Griest,
  Phys.\ Rept.\  {\bf 267}, 195 (1996)
  [hep-ph/9506380].

\bibitem{Su:2010qj} 
  M.~Su, T.~R.~Slatyer and D.~P.~Finkbeiner,
  Astrophys.\ J.\  {\bf 724}, 1044 (2010)
  [arXiv:1005.5480 [astro-ph.HE]].

\bibitem{Hooper:2010mq} 
  D.~Hooper and L.~Goodenough,
  Phys.\ Lett.\ B {\bf 697}, 412 (2011)
  [arXiv:1010.2752 [hep-ph]],
  D.~Hooper and T.~R.~Slatyer,
  Phys.\ Dark Univ.\  {\bf 2}, 118 (2013)
  [arXiv:1302.6589 [astro-ph.HE]].
  
\bibitem{Adriani:2008zr} 
  O.~Adriani {\it et al.}  [PAMELA Collaboration],
  Nature {\bf 458}, 607 (2009)
  [arXiv:0810.4995 [astro-ph]].

\bibitem{Aguilar:2013qda} 
  M.~Aguilar {\it et al.}  [AMS Collaboration],
  Phys.\ Rev.\ Lett.\  {\bf 110}, 141102 (2013).

\bibitem{Akerib:2013tjd} 
  D.~S.~Akerib {\it et al.}  [LUX Collaboration],
  arXiv:1310.8214 [astro-ph.CO].

\bibitem{Cheng:2002ej} 
  H.~C.~Cheng, J.~L.~Feng and K.~T.~Matchev,
  Phys.\ Rev.\ Lett.\  {\bf 89}, 211301 (2002)
  [hep-ph/0207125].

\bibitem{Servant:2002aq} 
  G.~Servant and T.~M.~P.~Tait,
  Nucl.\ Phys.\ B {\bf 650}, 391 (2003)
  [hep-ph/0206071].
  

\bibitem{Servant:2002hb} 
  G.~Servant and T.~M.~P.~Tait,
  New J.\ Phys.\  {\bf 4}, 99 (2002)
  [hep-ph/0209262].

\bibitem{Arrenberg:2008wy} 
  S.~Arrenberg, L.~Baudis, K.~Kong, K.~T.~Matchev and J.~Yoo,
  Phys.\ Rev.\ D {\bf 78}, 056002 (2008)
  [arXiv:0805.4210 [hep-ph]].

\bibitem{Cheng:2003ju} 
  H.~-C.~Cheng and I.~Low,
  JHEP {\bf 0309}, 051 (2003)
  [hep-ph/0308199].
  
\bibitem{Cheng:2004yc} 
  H.~-C.~Cheng and I.~Low,
  JHEP {\bf 0408}, 061 (2004)
  [hep-ph/0405243].

\bibitem{Low:2004xc} 
  I.~Low,
  JHEP {\bf 0410}, 067 (2004)
  [hep-ph/0409025].

\bibitem{Hubisz:2004ft} 
  J.~Hubisz and P.~Meade,
  Phys.\ Rev.\ D {\bf 71}, 035016 (2005).
  
\bibitem{Birkedal:2006fz} 
  A.~Birkedal, A.~Noble, M.~Perelstein and A.~Spray,
  Phys.\ Rev.\ D {\bf 74}, 035002 (2006)
  [hep-ph/0603077].

\bibitem{Asano:2006nr} 
  M.~Asano, S.~Matsumoto, N.~Okada and Y.~Okada,
  Phys.\ Rev.\ D {\bf 75}, 063506 (2007)
  [hep-ph/0602157].
 

\bibitem{Kanemura:2010sh} 
  S.~Kanemura, S.~Matsumoto, T.~Nabeshima and N.~Okada,
  Phys.\ Rev.\ D {\bf 82}, 055026 (2010)
  [arXiv:1005.5651 [hep-ph]].

\bibitem{Lebedev:2011iq} 
  O.~Lebedev, H.~M.~Lee and Y.~Mambrini,
  Phys.\ Lett.\ B {\bf 707}, 570 (2012)
  [arXiv:1111.4482 [hep-ph]].

\bibitem{Djouadi:2011aa} 
  A.~Djouadi, O.~Lebedev, Y.~Mambrini and J.~Quevillon,
  Phys.\ Lett.\ B {\bf 709}, 65 (2012)
  [arXiv:1112.3299 [hep-ph]].
  
\bibitem{Farzan:2012hh} 
  Y.~Farzan and A.~R.~Akbarieh,
  JCAP {\bf 1210}, 026 (2012)
  [arXiv:1207.4272 [hep-ph]].
 
\bibitem{Baek:2012se} 
  S.~Baek, P.~Ko, W.~I.~Park and E.~Senaha,
  JHEP {\bf 1305}, 036 (2013)
  [arXiv:1212.2131 [hep-ph]].
 
\bibitem{Yu:2014pra} 
  J.~H.~Yu,
  arXiv:1409.3227 [hep-ph].

 
\bibitem{Gasser:1990ce} 
  J.~Gasser, H.~Leutwyler and M.~E.~Sainio,
  Phys.\ Lett.\ B {\bf 253}, 252 (1991).

\bibitem{Mallot:1999qb} 
  G.~K.~Mallot,
  Int.\ J.\ Mod.\ Phys.\ A {\bf 15S1}, 521 (2000)
  [eConf C {\bf 990809}, 521 (2000)]
  [hep-ex/9912040].
  
\bibitem{Aprile:2012nq} 
  E.~Aprile {\it et al.}  [XENON100 Collaboration],
  Phys.\ Rev.\ Lett.\  {\bf 109}, 181301 (2012)
  [arXiv:1207.5988 [astro-ph.CO]].
  
\bibitem{Aprile:2012zx} 
  E.~Aprile [XENON1T Collaboration],
  arXiv:1206.6288 [astro-ph.IM].

\bibitem{xenon10T} 
Xenon10T, G3 expected sensitivity (2013), SNOMASS on the Mississippi (CSS 2013), SLAC Workshop Talk, XENON
http://www.snowmass2013.org/tiki-index.php?page=XENON; DMtools http://dmtools.brown.edu.


\bibitem{SIMPLE} 
M. Felizardo {\it et al.}   [The SIMPLE Collaboration],
  arXiv:1106.3014 [astro-ph.CO]. Phys.\ Rev. \ Lett. {\bf 108}, 201302, (2012)


\bibitem{COUPP} 
E. Behnke   {\it et al.}   [The COUPP Collaboration],
arXiv:1204.3094  [astro-ph.CO] Phys.\ Rev. \ D {\bf 86}, 052001,(2012)

\bibitem{Aprile:2013doa} 
  E.~Aprile {\it et al.}  [XENON100 Collaboration],
  Phys.\ Rev.\ Lett.\  {\bf 111}, no. 2, 021301 (2013)
  [arXiv:1301.6620 [astro-ph.CO]].


\end{thebibliography}
\end{document}